\documentclass[%
reprint,showpacs,nofootinbib,preprintnumbers,
 amsmath,amssymb,aps,
]{revtex4-1}

\usepackage{graphicx, fancybox, color}
\usepackage{amsmath,amssymb,bm}
\usepackage{braket}

\begin{document}

\title{
Issues with Search for Critical Point in QCD with Relativistic Heavy Ion Collisions
}

\author{Masayuki Asakawa}
\email{yuki@phys.sci.osaka-u.ac.jp}
\affiliation{
 Department of Physics, Osaka University, Toyonaka, Osaka 560-0043, Japan
}
\author{Masakiyo Kitazawa}
\email{kitazawa@phys.sci.osaka-u.ac.jp}

\affiliation{
 Department of Physics, Osaka University, Toyonaka, Osaka 560-0043, Japan
}
\affiliation{J-PARC Branch, KEK Theory Center, Institute of Particle and Nuclear Studies, 
KEK, 203-1 Shirakata, Tokai, Ibaraki 319-1106, Japan
}
\author{Berndt M\"uller}
\email{mueller@phy.duke.edu}
\affiliation{
Department of Physics, Duke University, Durham, NC 27708-0305, U.S.A.
}

\date{\today}

\begin{abstract}
A systematic search for a critical point in the phase diagram of QCD matter 
is underway at the Relativistic Heavy Ion Collider (RHIC) and is planned at
several future facilities. Its existence, if confirmed, and its location will greatly
enhance our understanding of QCD. In this note we emphasize several important 
issues that are often not fully recognized in theoretical interpretations of 
experimental results relevant to the critical point search. We discuss ways
in which our understanding on these issues can be improved.
\end{abstract}

\preprint{J-PARC-TH-0208}

\maketitle

Considerable experimental efforts have been made or are underway and
several new facilities are being planned to search for
a critical point in the QCD phase diagram with relativistic heavy ion collisions.
If the existence and location of a critical point can be ascertained,
it will become the second confirmed ``point'' in the QCD phase diagram
that has a physical meaning, after the ground state of nuclear matter. 
It will act as a landmark in the QCD phase diagram, and 
our understanding of the phases of QCD will be much deepened.
This is the reason why the critical point search is so important and
attracts great scientific interest.

In order to understand the effects of the existence of
a critical point on physical observables, it is necessary
to understand the roles and features of each observable. In the search of
critical point, fluctuations of conserved charges
\cite{Asakawa:2000wh,Jeon:2000wg,Bazavov:2012jq,Bazavov:2017tot} are most frequently
studied. However, as we will discuss in the following, the important
feature of fluctuations of conserved charges, which motivated the
original proposal, fluctuations of conserved charges as detectors
of the change of the degrees of freedom \cite{Asakawa:2000wh,Jeon:2000wg}, is often forgotten.
As a result, conserved charge fluctuations, or more precisely their cumulants, are often not
properly utilized in the current interpretation of experimental
results. In this short note, we discuss how we should understand
the experimentally observed conserved charge cumulants, focusing
on three issues.

Firstly, we argue at which freeze-out, chemical or kinetic, conserved
charge cumulants decouple. Very often, conserved charge cumulants
are assumed to decouple at the chemical freeze-out surface. 
However, this contradicts the notion of chemical freeze-out.
Chemical freeze-out is where most inelastic reactions cease,
i.e. where the changes of particle numbers and particle species
stop. Examples of such reactions are $\pi\pi \leftrightarrow \pi\pi\pi\pi$ and 
reactions that change the number of strange and antistrange valence quarks,
such as $\pi\pi \leftrightarrow K\bar{K}$ or $N\pi \leftrightarrow \Lambda K$.
(Note that inelastic reactions which do not require much energy
and have large cross sections, 
such as $n\pi^+ \leftrightarrow p\pi^0$ via the $\Delta^+$ resonance
and $p\bar{p} \leftrightarrow 3\pi$ are exceptions
\cite{Kitazawa:2011wh,Kitazawa:2012at}.) Observables involving conserved
charges are ``frozen'' in this sense at all times as they cannot be changed by local processes.
However, as emphasized in \cite{Asakawa:2000wh,Jeon:2000wg,Shuryak:2000pd},
fluctuations of conserved charges
change through diffusion, which does not require inelastic
scatterings and thus persists until kinetic freeze-out
(In ref.~\cite{Albright:2015uua} Albright et al. observed that the crossover equation of state can reproduce the data if the fluctuations are frozen at a temperature significantly lower than the average chemical freeze-out.).
The change of fluctuations of conserved charges does not proceed by
chemical reactions, because chemical reactions do not
change the net charge. It is diffusion that changes the fluctuations of
conserved charges. This was qualitatively discussed in \cite{Asakawa:2000wh}
and more quantitatively shown in \cite{Shuryak:2000pd}. Thus, it is necessary
to study the diffusion dynamics in the hadron phase between
the neighborhood of the critical point and the kinetic freeze-out
surface in order to explore the potential trace of the critical
point in the observed conserved charge fluctuations.

Secondly, we discuss the relationship between the cumulants
of conserved charges and the correlation length $\xi$ of the order
parameter field $\sigma$:
\begin{equation}
\langle \sigma({\mathbf x})\sigma({\mathbf y})\rangle - \langle\sigma\rangle^2
\propto e^{-|{\mathbf x}-{\mathbf y}|/\xi} .
\end{equation}
Stephanov \cite{Stephanov:2008qz} showed that in equilibrium
\begin{equation}
K_{n} \propto \xi^{m_n},
\label{cumulants}
\end{equation}
where $K_n$ is the $n$-th cumulant of a particle number to which the order parameter couples 
and $m_2\simeq 2,~m_3\simeq 4.5,~m_4\simeq 7,~m_5\simeq 9.5,~m_6\simeq 12$.
These relations, in particular, the increasingly large powers
for the higher cumulants, caught the attention of experimentalists
as they imply that higher cumulants are more sensitive
to the critical point. Led by this expectation, many experimental efforts
are being made to measure higher cumulants (fourth and sixth order)
of net protons as proxies of net baryon number cumulants.

Here we need to remember the following fact. 
In the case of conserved charge fluctuations,
the left-hand side of Eq.~(\ref{cumulants}) is a conserved
quantity; it can change only through diffusion, i.e.\ by particle transport.
Neither pair production nor pair annihilation changes its
value. On the other hand, the right-hand side of
Eq.~(\ref{cumulants}) is not a conserved quantity; 
it can change due to propagation of information related to the order parameter.
Thus the left-hand side and right-hand side obey different
equations of motion; if the conserved charges are
associated with quasi-particles, the left-hand side is achieved
only by particle transport, while the right-hand side can
be changed by propagation of, e.g. the information of
the amplitude of the order parameter field,
which obeys a wave equation (see, e.g. \cite{Paech:2003fe}).

In nonequilibrium or in dynamical situations
as in relativistic heavy ion collisions, there is no guarantee
that the proportionality relation (\ref{cumulants}), holds
(see, e.g. \cite{Kitazawa:2013bta,Nahrgang:2018afz}).
The left-hand side changes by diffusion and evolves only slowly. 
This observation motivated the proposal that fluctuations may be probes of
quark deconfinement \cite{Asakawa:2000wh,Jeon:2000wg}.
The change of the right-hand side is not constrained by particle diffusion 
but governed by the dynamics of the critical mode. 

As the evolution of a conserved quantity is much slower than 
that of the order parameter field $\sigma$, 
on longer time scales the order parameter field follows the
evolution of the conserved quantity and plays no independent role
in the dynamics \cite{Son:2004iv,Fujii:2004jt}.
Assuming this separation of the time scales, the evolution of the
conserved quantity near the critical point has been studied recently
\cite{Sakaida:2017rtj,Nahrgang:2018afz}.

Because of the finiteness of the reaction time and critical slowing down, 
however, the growth of the correlation length in 
relativistic heavy ion collisions is limited even if the system passes 
right through the critical point
\cite{Fujii:2003bz,Fujii:2004jt,
Berdnikov:1999ph,Nonaka:2004pg}.
In other words, even the fast mode, the order parameter field $\sigma$,
does not reach equilibrium; neither do quantities related
to the slow mode, the cumulants of the conserved quantity.
In order to confirm to what extent the relation \eqref{cumulants} holds
or is violated in relativistic heavy ion collisions, dynamical calculations
that treat the fast order parameter field and the slow hydrodynamical
fields properly, as described in \cite{Fujii:2003bz,Fujii:2004jt} 
for the static case, will be needed.

To summarize, Eq.~(\ref{cumulants}) does
not hold in general in relativistic heavy ion collisions.
Combining this observation and the argument in the previous
paragraph, we conclude that one needs to follow the evolution
of cumulants from the quark phase down to kinetic freeze-out
by properly taking account of the diffusive property of conserved 
charge fluctuations without resort to relation (\ref{cumulants}) 
in order to compare theoretical expectations
and experimental results.

Finally, we need to recognize that experimentally observed cumulants
are measured in momentum space with an acceptance, e.g, in the
case of STAR $|\eta| \leq 1.6$, while theoretical cumulants
are usually calculated in coordinate space. In order to compare
experimental results and theoretical calculations, one needs
a map from coordinate space to momentum space. Only in limited cases, 
such as for a one-dimensional boost-invariant (Bjorken) expansion,
the result in coordinate space is identical to the one in momentum
space. Note that even in this case, the identity holds only up to thermal
smearing \cite{Ohnishi:2016bdf}.
For this purpose, i.e. to have the map, construction
of a dynamical model of the reaction is inevitable. In the energy
range of Beam Energy Scan (I\hspace{-.1em}I), obviously the Bjorken picture
is not appropriate. It may be worthwhile considering the case
of the Landau-Fermi picture as an opposite extreme to the Bjorken
scenario. 

To sum up, the collision geometry and its time evolution
are necessary to understand the final state fluctuations, which
can be compared with experimental results. Note that it is not
sufficient to know only the chemical and kinetic freeze-out surfaces, 
since the evolution of conserved charge fluctuations should be traced
all the way from the initial state to kinetic freeze-out as
we discussed above.  These three issues are often not adequately 
taken into account in the interpretation
of the experimental results for conserved charge cumulants.

At the end of this note, we point out two related issues.
One concerns the net proton number cumulants. These are often 
considered as proxies for the net baryon number cumulants which 
are conserved quantities. As we discussed above
and as Refs.~\cite{Kitazawa:2011wh,Kitazawa:2012at}
pointed out, net proton number cumulants are not conserved
locally in the hadron phase because they can change
through reactions such as $p\pi^- \rightarrow n\pi^0$.
These reactions do not require much energy and continue to occur 
after what is commonly called chemical freeze-out.
The same argument should also be applied when one regards the net kaon
number cumulants as proxies of the cumulants of net strangeness.

The other is the initial condition in the Landau-Fermi
picture. At the collision energies that produce near-critical
point QCD matter, the transverse correlation length of nuclear
energy density is probably best estimated
using the wounded nucleon model, implying a correlation
length comparable to the nucleon radius, approximately 1 fm. 
This correlation evolves while the two nuclei collide, overlap, and
eventually stop. When the two nuclei have stopped, 
the typical correlation length in the transverse
direction is estimated to be of order $2R/c_s \gamma$.
Here $R$ is the radius of the colliding nuclei and $c_s \approx 0.4$ 
is the speed of sound in baryon rich QCD matter.
This correlation could be realized in simulations
as correlated domains whose transverse
size is of the order of $2R/c_s \gamma$ with a finite transverse
flow gradient in the hydrodynamic initial conditions. This would
give an improved initial condition for the Landau-Fermi picture
implementing baryon stopping.

We thank Swagato Mukherjee (Brookhaven National Laboratory (BNL))
for discussion of the topics discussed in this manuscript.
M.A. thanks BNL for support during his visit. M.A. and M.K. acknowledge support from JSPS KAKENHI Grants No. 195
JP18K03646, No. JP17K05442 and No. JP19H05598, respectively.
B.M. acknowledges support from U.S. Department of Energy 
grant No. DE-FG02-05ER41367.


\begin{thebibliography}{99}

\bibitem{Asakawa:2000wh} 
  M.~Asakawa, U.~W.~Heinz, and B.~M\"uller,
  Phys.\ Rev.\ Lett.\  {\bf 85}, 2072 (2000).
\bibitem{Jeon:2000wg} 
  S.~Jeon and V.~Koch,
  Phys.\ Rev.\ Lett.\  {\bf 85}, 2076 (2000).
\bibitem{Bazavov:2012jq} 
  A.~Bazavov {\it et al.} [HotQCD Collaboration],
   Phys.\ Rev.\ D {\bf 86}, 034509 (2012).
\bibitem{Bazavov:2017tot} 
  A.~Bazavov {\it et al.} [HotQCD Collaboration],
  Phys.\ Rev.\ D {\bf 96}, no. 7, 074510 (2017).
\bibitem{Kitazawa:2011wh} 
  M.~Kitazawa and M.~Asakawa,
  Phys.\ Rev.\ C {\bf 85}, 021901 (2012).
\bibitem{Kitazawa:2012at} 
  M.~Kitazawa and M.~Asakawa,
  Phys.\ Rev.\ C {\bf 86}, 024904 (2012)
  Erratum: [Phys.\ Rev.\ C {\bf 86}, 069902 (2012)].
\bibitem{Shuryak:2000pd} 
  E.~V.~Shuryak and M.~A.~Stephanov,
  Phys.\ Rev.\ C {\bf 63}, 064903 (2001).
\bibitem{Albright:2015uua} 
  M.~Albright, J.~Kapusta and C.~Young,
  Phys.\ Rev.\ C {\bf 92}, no. 4, 044904 (2015).
\bibitem{Stephanov:2008qz} 
  M.~A.~Stephanov,
  Phys.\ Rev.\ Lett.\  {\bf 102}, 032301 (2009).
\bibitem{Paech:2003fe} 
  K.~Paech, H.~St\"ocker, and A.~Dumitru,
  Phys.\ Rev.\ C {\bf 68}, 044907 (2003).
\bibitem{Kitazawa:2013bta} 
  M.~Kitazawa, M.~Asakawa, and H.~Ono,
  Phys.\ Lett.\ B {\bf 728}, 386 (2014).
\bibitem{Nahrgang:2018afz} 
  M.~Nahrgang, M.~Bluhm, T.~Sch\"afer, and S.~A.~Bass,
  Phys.\ Rev.\ D {\bf 99}, no. 11, 116015 (2019).
\bibitem{Son:2004iv} 
  D.~T.~Son and M.~A.~Stephanov,
  Phys.\ Rev.\ D {\bf 70}, 056001 (2004).
\bibitem{Fujii:2004jt} 
  H.~Fujii and M.~Ohtani,
  Phys.\ Rev.\ D {\bf 70}, 014016 (2004).  
\bibitem{Sakaida:2017rtj} 
  M.~Sakaida, M.~Asakawa, H.~Fujii, and M.~Kitazawa,
  Phys.\ Rev.\ C {\bf 95}, no. 6, 064905 (2017).
\bibitem{Berdnikov:1999ph} 
  B.~Berdnikov and K.~Rajagopal,
  Phys.\ Rev.\ D {\bf 61}, 105017 (2000).
\bibitem{Fujii:2003bz} 
  H.~Fujii,
  Phys.\ Rev.\ D {\bf 67}, 094018 (2003).
\bibitem{Nonaka:2004pg} 
  C.~Nonaka and M.~Asakawa,
  Phys.\ Rev.\ C {\bf 71}, 044904 (2005).
\bibitem{Ohnishi:2016bdf} 
  Y.~Ohnishi, M.~Kitazawa, and M.~Asakawa,
  Phys.\ Rev.\ C {\bf 94}, no. 4, 044905 (2016).

\end{thebibliography}
\end{document}